# Crystallization Mechanism Tuned Phase-Change Materials: Quantum Effect on Te-Terminated Boundary

*Wen-Xiong Song[1#]\*, Qiongyan Tang[#2], Jin Zhao[#1,3], Muriel Veron[4], Xilin Zhou[1], Yonghui Zheng[2], Daolin Cai[1], Yan Cheng[2]\*, Tianjiao Xin[1], Zhi-Pan Liu[5], and Zhitang Song[1]\**

[1]State Key Laboratory of Functional Materials for Informatics, Shanghai Institute of Microsystem and Information, Chinese Academy of Sciences, Shanghai 200050, China; [2]Key Laboratory of Polar Materials and Devices (MOE), Department of Electronics, East China Normal University, Shanghai 200241, China; [3]School of Physical Science and Technology, Shanghai Tech University, Shanghai 201210, China; [4]University Grenoble Alpes, CNRS, SIMAP, 38000 Grenoble, France; [5]Department of Chemistry, Fudan University, Shanghai 200433, China.

While phase-change materials (PCMs) composed of chalcogenide have different crystallization mechanisms (CM), such as nucleation-dominated $Ge_2Sb_2Te_5$ (GST) and growth-dominated GeTe (GT), revealing the essential reason of CM as well as the tuned properties is still a long-standing issue. Here, we remarkably find the distinct stability of Te-terminated (111) boundaries (TTB) in different systems, which provides a path to understand the difference in CM. It stems from the quantum effect of molecular orbital theory: the optimal local chemical composition results in the formation of TTB without dangling bonds (DB) in GST but with DB in GT, where DB destabilizes boundary due to its distorted local environment mismatching $O_h$ symmetry of $p$ orbitals. Moreover, the inner vacancy concentration in GST is alterable and controlled by TTB, manifested by the absence of cubic-to-hexagonal transition in carbon-doped GST of small grains and minimized inner vacancy. Finally, the charge transport property (CTP) is controlled by boundary via changing the density of charge or hole nearby as well as vacancy. These findings open the door to tune CTP by CM, which is necessary for achieving low-power and ultrafast devices.

## Introduction

Grain refinement (GR) is a general approach to alter material properties [1] by introducing planar defects, which influence not only conventional mechanical property [2] but also thermal [3] and electrical [4] transport properties, such as electronic structure of tuning band gap [5] and catalytic properties [6]. It is not exception in the field of phase-change random access memory (PCRAM) as one of the most mature emerging nonvolatile memory (eNVM) technologies [7] for the applications of processing-in-memory and neuro-inspired computing technologies [8], which utilizes a fast (~ns) and reversible amorphous-to-crystal transition in phase-change materials (PCMs) [9,10]. It is noted that current cell size is confined to ~20 nm in width and ~40 nm in length [11], which can be further shrunk [12] for higher density storage and much lower power consumption [13]. The improved propformance has been raveled by GR engineering through controlling the thermal and mass transport properties. [3] However, different PCMs, such as $(GeTe)_{1-x}–(Sb_2Te_3)_x$ pseudobinary alloys and $Sb_2Te$, present much distinct grain size. The size of $Ge_2Sb_2Te_5$ (GST) changes from several nanometers near crystallization temperature ($T_c$, 423 K) to abnormal small size of 10~30 nm at 523 K, albeit higher than the half melting temperature ($T_m$, 900K). By contrast, the size of both GeTe (GT) and $Sb_2Te$ is much larger of hundreds of nanometers at their corresponding $T_c$. [14] These two categories of PCMs are classified as nucleation- and growth-dominated crystallization, which is the result of speed competition between nucleation rate and grain boundary (GB) migration but with few understanding.

[15] It is thus a pressing need and challenge to uncover the different GB stability in various PCMs for better achieving GR and tuning functional properties.

GR is always achieved by general thermodynamic or kinetic strategies [16], where the former reduces GB energy [17] and the latter decreases GB mobility via various ways, such as solute drag [16] and second-phase drag [18]. From the thermodynamic perspective, the surfaces of PCMs, a little different from GB, have been systematically studied in the experiment [19,20] and theory [21,22], which has manifested that Te-terminated surface is energetically favorable in both GeTe and Ge−Sb−Te systems. However, their different crystallization mechanisms indicate their distinct GB stability, even for the same preferred Te-terminated boundary (TTB).

From the kinetic perspective, some impurities replace lattice positions and act as solute drag, such as Sc [23] and Ti [24]; some nonmetal impurities gather in the boundary and act as a second phase, such as carbon [3,25]. It is found that GT changes to nucleation domination after doping Sc (SGT). [14] The transition also occurs from the growth-dominated $Sb_2Te$ to nucleation domination after doping $Al_3Sc$ alloy (ASST). [26] It indicates that GST has an ignored factor that can stabilize GB. The obvious difference between GST and GT is vacancy, where there are 20% vacancies in cationic positions of GST compared with no vacancy in ideal GT. The introduced vacancy, whose content is $x/(2+4x)$ for $(GeTe)_{1-x}–(Sb_2Te_3)_x$ [27], stabilizes crystalline phases by reducing antibonding sates [28]. Antibonding analysis was also investigated in amorphous [29,30]. These give us a hint that the existence of certain interface or local structure stabilizes grains in GST from the quantum effect.

On the other hand, while vacancy has attracted much attention because of its effect on metal-insulator transition [31,32], how vacancy produces still puzzles us [33,34]. The homogenous crystallization result shown no inner vacancy in small grains [33], but the crystal-growth model observed inner vacancy during the boundary migration [34]. Grain-size correlated boundary is a key to understand this difficulty.

To shed light on these issues, we at first analyze the distinct stability of TTBs in various PCMs using molecular orbital (MO) theory combined with first principle calculations. Experimentally, to verify the prediction of MO theory, we compare the GB features of the crystal-crystal and crytal-amorphous interfaces in both GST and GT. To manifest fewer vacancies in small grains, we study C-doped GST (C−GST), which has excellent device performance [3,25,35,36]. Theoretically, we study the boundary porperties of general stability and charge transport property (CTP).

## Results

**MO explains the stability of TTB.** Our work starts by investigating the stability of TTB in PCMs using the molecular orbital (MO) theory. MO theory uses a linear combination of atomic orbitals to represent molecular orbitals resulting from bonds between atoms, which are divided into bonding, antibonding, and non-bonding types. Figures 1(a)-(c) show the schematic MO diagram for Sb−Te, Ge−Te, and Ge−Sb−Te systems, respectively. Here, we only consider the interactions among $p$ orbitals, because of full filled $s$ orbitals of deep energy levels far from $p$ orbitals and hard to be hybridization.

We find the different optimal cation/anion ratio for the three systems as their outermost orbitals are full filled. The optimal ratio is $Sb_2Te_3$ (ST), GeTe (GT), and $(GeTe)_{1-x}$−$(Sb_2Te_3)_x$, such as $Ge_2Sb_2Te_5$ (GST). It is noted that ST exists lone-pair electrons and forms non-bonding. In GT, the lone-pair $p$ electrons of Te atoms fills the empty $p$ orbital of Ge atom, which makes Ge−Te interaction stronger than Sb−Te interaction. Because of equal cationic and anionic positions in rock-salt structure, it is suggested that the content of random vacancy is 16.67% for ST, zero for GT, and 10% for GST, which agrees with experimental result [27].

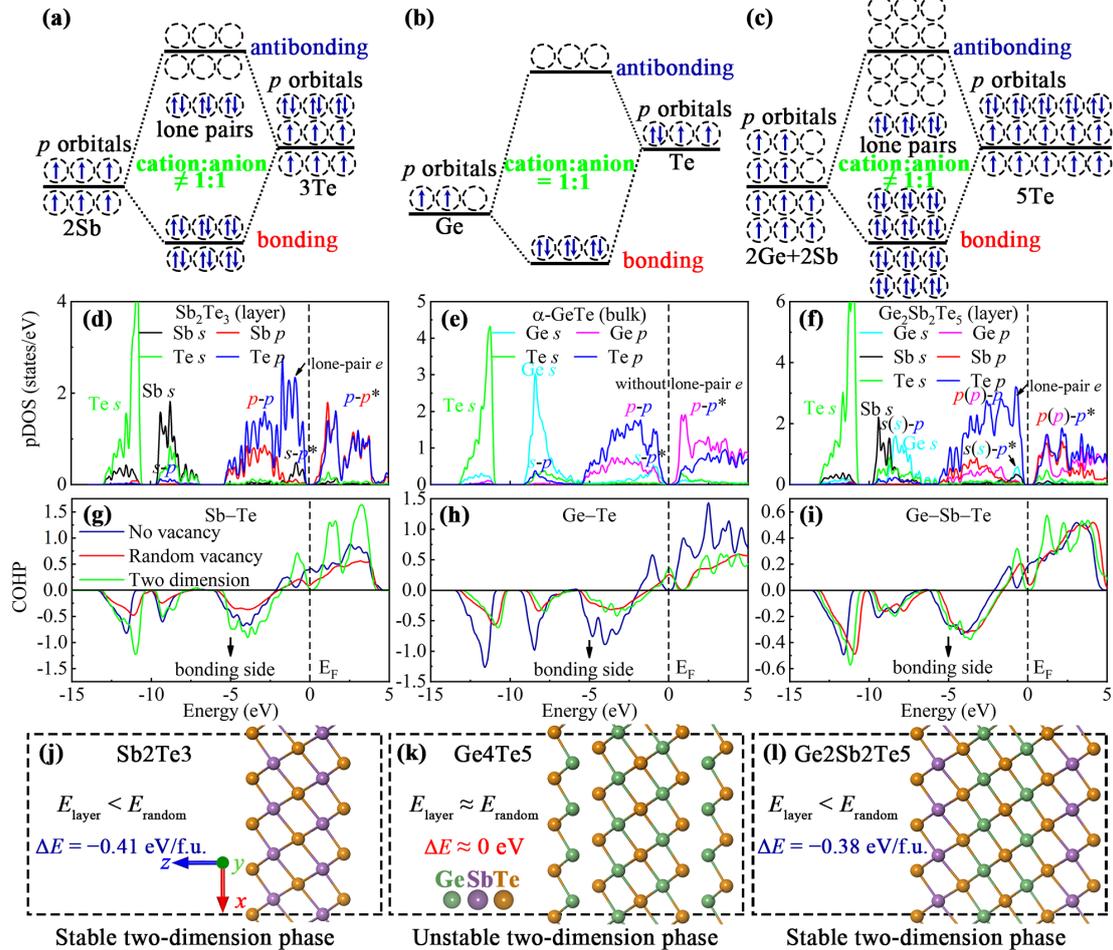

**Figure 1.** MO theory explains the stability of exposed TTB judged by whether forming full filled outermost orbitals. **(a)-(c)**, the schematic MO diagram for Sb−Te **(a)**, Ge−Te **(b)**, and Ge−Sb−Te **(c)**. **(d)-(i)**, evidences of orbital interactions by calculating the pDOSs and COHP. Figure **(g)-(i)** show the average COHP of all interactions within the cutoff of 3.6 Å in the no-vacancy model (bulk, blue), random-vacancy model (bulk, red), and two-dimension phases with exposed TTB (green). **(j)-(l)**, the relative stability of the phase with exposed TTB compared with the phase with random vacancy. Whether the optimal cation/anion ratio is one or not determines the stability of TTBs: it is stable in two-dimension $Sb_2Te_3$ and $Ge_2Sb_2Te_5$ phases, while it is unstable in two-dimension Ge−Te (formula $Ge_4Te_5$).

To verify the MO theory, we analyze the partial density of states (pDOS) and crystal orbital Hamilton population (COHP), as shown in Fig. 1(d)-(i). Figure 1(d)-(f) show all the bonding $p–p$ orbitals below Fermi level and all the anti-bonding $p–p*$ orbitals above Fermi level. We find the existence of lone-pair electrons as non-bonding interaction just below Fermi level in both ST and GST, but not existence in GT. All the $s$ orbitals mainly locate at deep energy levels, albeit a weak hybridization between the $s$ orbitals (Ge or Sb) with $p$ orbitals (Te), where the anti-bonding $s–p*$ orbitals of the three systems are just below Fermi level.

Figures 1(g)-(i) emphasize the importance of vacancy (or TTB) from COHP analysis. Three models, i.e. no vacancy, random vacancy, and two-dimension model, are constructed. The formula of no vacancy model are SbTe, GeTe, and $Ge_2Sb_2Te_4$ and the random vacancy models are $Sb_2Te_3$ (16.67% vac.), $Ge_4Te_5$ (10% vac.), and $Ge_2Sb_2Te_5$ (10% vac.). For the no vacancy model, figures 1(g)-(i) show that the fermi level of GT is at the bonding state, but ST and GST locate at antibonding states, originating from the redundant electrons occupy the anti-bonding $p–p*$ orbitals. On the contrary, for the random vacancy and two dimension models, the $p–p$ antibonding interactions in GST and ST systems remove; the $p–p$ bonding interaction in GT become weaker, albeit the reduced anti-bonding $s-p*$ interaction. The COHP calculations demonstrate that vacancy (or TTB) stabilizes both ST and GST, which have been mentioned in the reference [28], and destabilzes GT. The essential reason of producing vacancy (or TTB) originates from its ability to tune the composition.

As vacancy gathering, we further investigate the stability of TTB comparing with the random vacancy model, as shown in Fig. 1(j)-(l). In ST and GST, their two-dimension layers are more stable than the corresponding vacancy random models, which may be caused by the better evironment for hybridization and in line with other calculations [31,37]. On the contrary, we have proved that vacancy destabilizes GT system, albeit the stability of the two-dimension model similar to the random vacancy model. A conclusion is made that TTB in both ST and GST is more stable than that in GT. In the following experiments, based on TTB, we reveal the reason of distinct crystallization mechanisms and study alterable inner vacancy concentration in small grains, both of which have significant effects on CTP.

**Observed stable TTB in GST.** Figure 2(a) shows the orientation map of crystallized GST film (200°C for 30min, $0.53T_m$). The distribution (in area) of grain size is mainly around 15~30 nm, as shown in Fig. 2(b), and increases a little at 250°C ($0.58T_m$, Fig. S1). Normally, the grain size in pure metals is larger than 100 nm above $0.5T_m$, albeit small grains in some reported alloys for the solute segregation in GBs. [16] Thus, we are confused with such small grain in GST because of no element segregation. [38]

Figure 2(e) shows that the GB of a [110] oriented cubic-phase grain is composed of TTB (111) plane (corresponding to the (0001) plane in hexagonal phase [19]), marked by white dash line, proving the extraordinarily stable TTB (111) in GST. A Σ3 (111) twin boundary (TB) with a vdW gap, named as vdW-TB, is found, as shown in Fig. 2(e) and enlarged in Fig. 2(f), using the atomic-resolution STEM high angle annular dark field (HAADF) images. We map out the positions of Te and Ge/Sb/vacancy because the contrast of HAADF image is approximately proportional to the square of atomic number ($Z = 6, 32, 51$, and $52$ for C, Ge, Sb and Te, respectively). The interval change of distance between Ge/Sb plane (white line) and Te plane (yellow line) is investigated in Fig. 2(g), manifesting the existence of Peierls distortion [28]. The 3.54 Å or 3.52 Å are consistent with the (111) interplanar distance in cubic phase, and the 2.82 Å is equal to the vdW gap. [38]

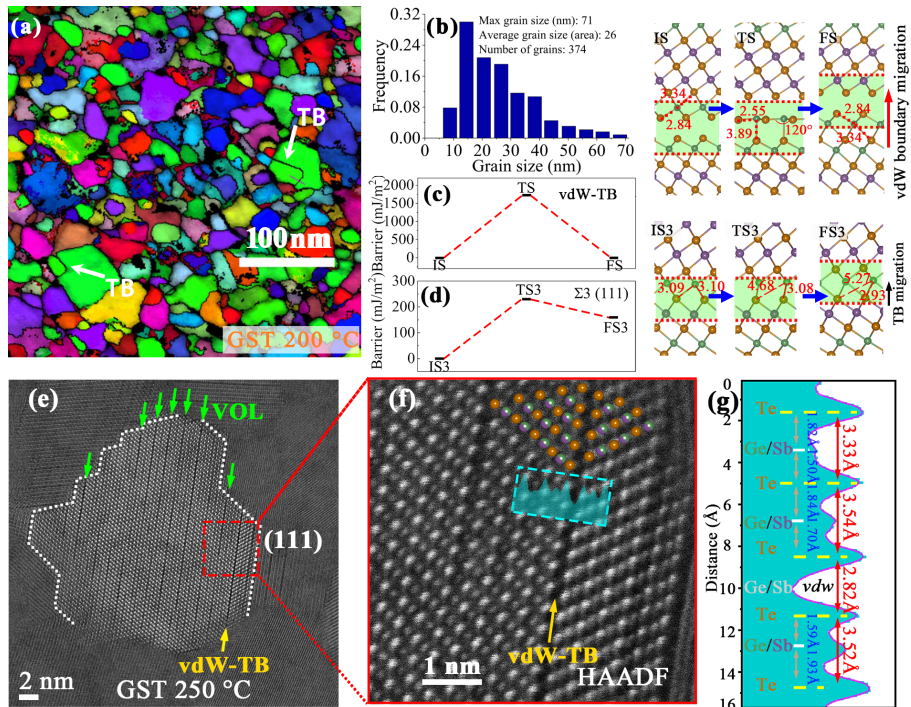

**Figure 2.** GB characteristics in GST. **(a)**, a crystal orientation map reconstructed from ACOM-TEM annealed at 200 °C for half an hour. **(b)**, grain size distribution. **(d)-(e)**, the migration barrier of vdW and Σ3 boundaries, whose side views of key intermediates states are shown in the top-right. **(f)-(g)**, STEM-HAADF images of TB boundary for **(f)** overall view and **(g)** local enlarged region after annealing 15 minutes at 250 °C. In **(f)**, the (111) plane is TTB; the circled region is probable the beginning crystallization region of no inner vacancy. **(h)**, the intensity profile of marked rectangle region in **(g)**.

It is noted that TBs found in large grains, but few TB observed in small grains, as shown in Fig. 2(a). It suggests that TB is formed during the crystal growth, such as via the migration of Σ3 TB, whose migration barrier is much low, as shown in Fig. 2(d). It will be stabilized by vacancy ordering layers (VOLs) and finally form a vdW-TB constituted of two TTBs, because high gap-migration barrier in Fig. 2(c) as well as its thermodynamic stability accroding to MO theory. Gap stacking faults are also found in other PCMs. [39]

**Crystallization mechanism determined by TTB stability.** In the following, we reveal the essential reason why different crystallization mechanisms exist in PCMs. It is empirically believed that a growth-dominated system has much lower nucleation rate. However, figure 3(a) shows that GST and GT exhibit approximate nucleation rates, according to the formula $J = 1/V\tau_J$, where $V$ is the volume of simulation box and $\tau_J$ is the crystallization time. According to the crystallization time of GST and GT at 600K as shown in Fig. 3a, to our surprise of approximate equal, their nucleation rates are estimated as $2.4\times10^{35}$ m$^{-3}$s$^{-1}$ and $1.7\times10^{35}$ m$^{-3}$s$^{-1}$, respectively, agreeing with the estimation from the previous calculations [40,41]. Thus, the determining reason may be their different boundary stability other than the nucleation rate.

Figure 2(b) shows a Σ3 (111) TB as well as a TB disappearing or forming region circled by orange in GT. Different from GST, no vdW gap is observed. TB in GT is thus destabilized, because the distorted local environment mismatches $O_h$ symmetry of $p$ orbitals. Figure 3(c) and 3(d) show the TEM images of GST and GT at $T_c$. Albeit their similar nucleation rates, the isolated crystalline grain in GST is about 10 nm; on the contrary, in GT, the grain size is over 100 nm (see Fig. S2), and grows up very rapidly as nucleation is completed. Figure 3(e) and 3(f) illustrate the typical boundaries of GST and GT, respectively, in which the boundaries in GST are all (111) planes, while the boundaries in GT are (111) and (110) planes, suggesting that TTB in GST is more stable than in GT, in line with our previous MO analysis. It should be mentioned that Sb$_2$Te is also growth domination [42], which has not enough TTB [43]. On the other hand, the impurity-induced growth-to-nucleation transition [14,26] originates from the kinetic constraint around dopant [40].

**Inner vacancy concentration tuned by TTB in small grains.** Subsequently, we investigate the inner-vacancy concentration, because vacancy has a significant effect on CTP [31]. Based on the fact of lots of Te atoms as well as vacancies in TTB, we predict that the inner vacancy concentration is alterable, particularly in small grains. Applying a sphere grain model of the Te-terminated (111) interface, Fig. 4a shows the ratio, the number of boundary Te atoms divided by the total number, changing with grain size, where small grain has much high ratio. More details see SI. We can predict that GST exists a critical grain size of 4.7 nm of no inner vacancy when the ratio is 11.11%. For ST, the critical size is 2.6 nm as the ratio is equal to 20%. The ratio will reduce as grain grows. In order to keep the optimal local composition, the vacancy should be produced in the inner, which is manfisted by the VOL in Fig. 2(e); meanwhile, occurrence of mainly nine-atom layers meet the optimal chemical composition. For the sample annealing at low temperature, 180 °C for half an hour, short VOLs are also produced shown in Fig. 4(h). The evidences of alterable vacancy concentration in grains of different size are provided by the references that no inner vacancy is in small grains [33] and inner vacancy is produced during the migration of an infinite boundary [34]. Small grains with low inner concentration prefer cubic phase, which is verified by doping carbon to refine the grains.

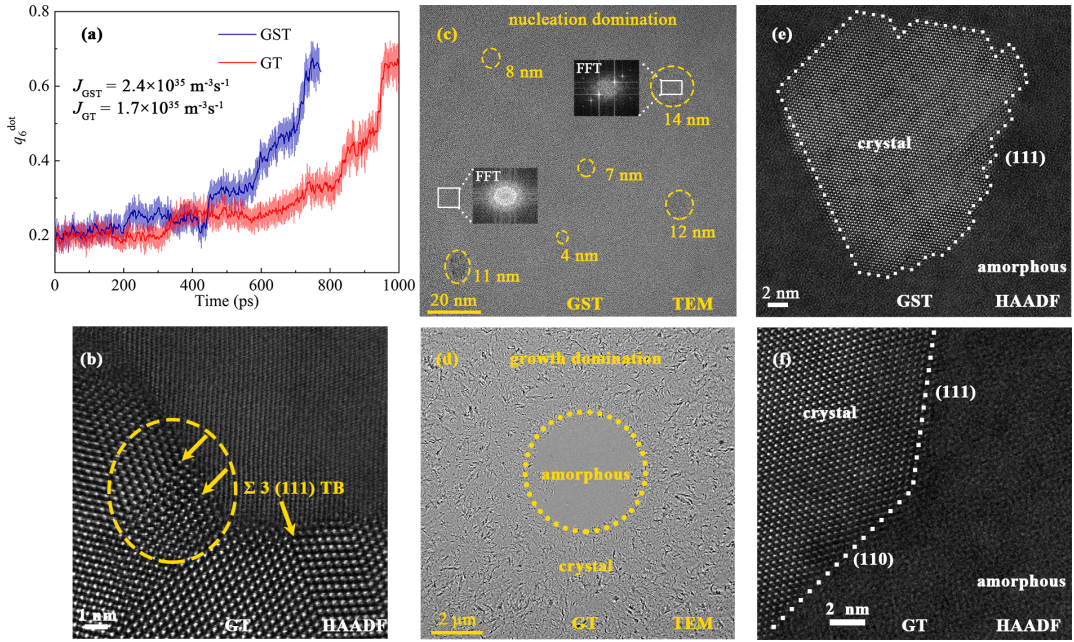

**Figure 3.** TTB stability determines the crystallization mechanism. **(a)**, similar nucleation rates at 600K for GST (blue) and GT (red) are deduced from their approximate cystallization time. **(b)**, a Σ3 (111) TB of no vdW gap and a TB disappearing region are observed in GT. **(c)** and **(d)** are the TEM images for nucleation-dominated GST and gowth-dominated GT, respectively. **(e)** and **(f)** are typical crystal-amorphous boundaries in GST and GT, respectively.

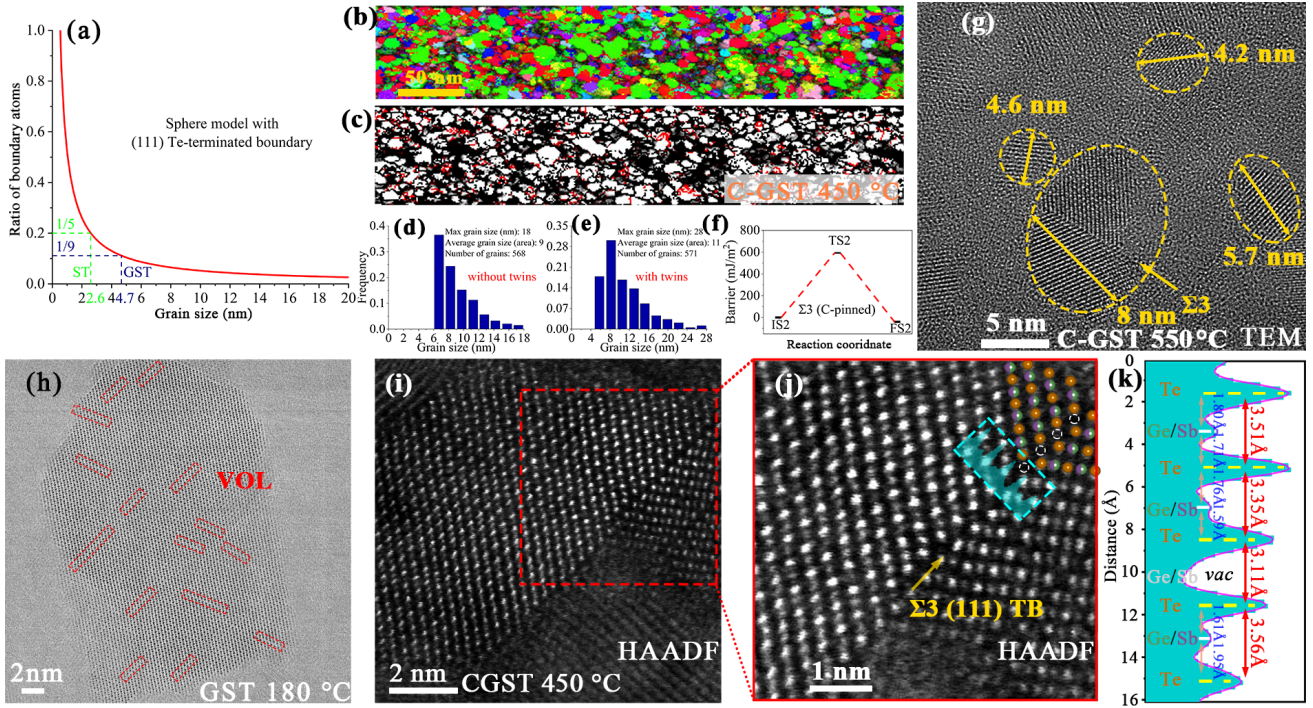

**Figure 4.** Evidences of iterable inner vacancy concentration. **(a)**, the relationship between the ratio of boundary atoms and grain size. **(b)**, crystal orientation maps reconstructed from ACOM-TEM for C−GST annealed at 450°C for 30min, where the red lines in **(c)** show the TB. Grain size distributions are shown without TB **(d)** and with TB **(e)**. **(f)**, the calculated barrier of Σ3 (111) TB with carbon in the boundary. **(g)**, the TEM figure of C−GST sample annealed at 550°C for 30min. **(h)**, observed VOLs in GST annealing at 180 °C for 30min. The STEM-HAADF images of TB boundary for **(i)** overall view of two grains and **(j)** local enlarged region in C−GST after annealing 30min at 450°C. **(k)**: The intensity profile of the marked rectangle region in **(j)**. The carmine line is the gaussian smearing result of cyan histograms.

Figure 4(b) shows the orientation color map of C−GST annealed half an hour at 450 °C. Compared with GST, the grain size dramatically reduces to several nanometers (~9 nm herein), whose distribution is shown in Fig. 4(d). Compared with crystalline GST with many VOLs, VOLs are much fewer in C−GST. All grains retain cubic structure even at 550 °C shown in Fig. 4(g), due to the small portion of vacancy (~1.6%) inside the grain, which is favorable in actual devices [44].

On the other hand, more TBs have formed in C−GST, whose boundaries are marked by red lines in Fig. 4(c). Regarding TB as new GB, the change of grain-size distribution, as shown in Fig. 4(e), predicts the content of TBs of more than 20%. Figure 4(i) and 4(j) show the STEM-HAADF image of one Σ3 (111) TB in C−GST. Figure 4(k) shows the intensity profile of the marked rectangle region in Fig. 4(j), simialr to the scenario in GST. It is noted that the 3.11 Å width between two Te planes in the boundary is larger than 2.82 Å. It illustrates the existence of atoms in the vacancy-aggregation layer and still a Σ3 (111) TB, suggesting that the invisible atomic images are probable C an Ge due to their smaller atomic number. To provide the evidence of TB stabilized by carbon, the migration barrier of kinetic constraint is 593 mJ/m$^2$, as shown in Fig. 4(f), which is higher than the shearing barrier of 230 mJ/m$^2$ in Σ3 (111) TB.

**Discussion**

**Distorted local environment destabilizes boundary**. In this section, we provide the evidence of distorted local environment destabilizing boundary. We focus on TB, which stores low enthalpy and refine grain by forming stable nanotwins [45-47], such as in Fe [48] and Cu [49]. However, the formation of TB in some alloys is difficult, such as Al [50], which has large generalized stacking fault energy (GSF) value than the previous two metals. [51] Figure 5(a) shows the GSF of the stacking and twin fault planar defects, the shapes of which determine the nucleation rate of forming twins or dislocation activity. The large $\gamma_{utf}/\gamma_{usf}$ value of 1.54 is similar to Al, Ni, Pt, and Pd, [51,52] with low nucleation rate, where the transition-state energy $\gamma_{utf}$ and $\gamma_{usf}$ are unstable twin fault energy and unstable stacking fault energy, respectively. The $\gamma_{sf}/\gamma_{usf}$ value of 0.79 is large and similar to Al [51], where $\gamma_{sf}$ is the stacking fault energy. It declares that TBs in PCMs is hard to form spontaneously in bulk, agreeing with above ACOM-TEM results of few TBs. A few fault stacking, such as the vdW-TB in Fig. 2, may be stabilized by VOL.

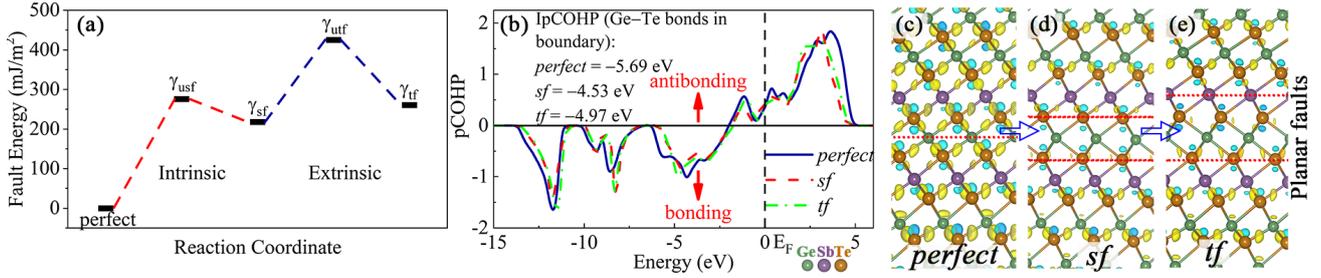

**Figure 5.** Evidences of distorted local environment destabilizes boundary. **(a):** A GSF curve of stacking faults obtained by shearing a perfect crystal along a (111) plane. **(b):** projected COHP (pCOHP) of the Ge–Te interactions in the bounary for the *perfect* (blue), *sf* (red), and *tf* (green), respectively. The integral pCOHP (IpCOHP) is shown in the inner. **(c)-(e):** CDD contours for the *perfect*, *sf*, and *tf* structures, where the 3D isosurface is ±0.047eÅ$^{-3}$. Yellow and cyan isosurfaces are positive and negative charge densities, respectively.

To explain the large GSF, we analyse the bonding by calculating the charge density differences (CDDs) of the *perfect*, *sf* and *tf* structures, as shown in Fig. 5(c)-(e). In the *perfect* structure, each atom has yellow isosurfaces, manifesting the existence of covalent bonds. But no yellow isosurface is found in the fault regions of both *sf* and *tf* structures. It illustrates weak bonds among them and the occurrence of electron transfer from the planar-fault region to others, which is opposite to the scenario of vacancy trapping electrons [31]. The pCOHP calculations in Fig. 5(b) manifest weaker bonds in planar faults. The increased interface energy is caused by atoms in the boundary with non-orthometic local environments mismatching the $O_h$ symmetry of $p$ orbitals, which is the reason of large GSF. Therefore, non-octahedral environments around GB destabilize its stability, which explains the large grain size in GT.

**Charge transport influenced by TTB.** In traditional semiconductors, planar defects often hinder charge transport. [53] Having proved the stable TTB in $(GeTe)_{1-x}-(Sb_2Te_3)_x$ ($x>0$), here we show the significant influence of boundary on electronic property. The band structures of two-dimensional structures are shown in Fig. 6(a)-(b), and their corresponding bulk strcutures with vdW gap are shown in Fig. 6(c)-(d), which are similar to the one considering spin-orbit coupling (SOC) [54]. All of them present Diract-cone-like. The two-dimensional structures present larger band gap, due to the quantum confinement and less interlayer coupling, which is also found in layered transition metal dichalcogenide (TMD) [55,56]. In both bulk and layer phases, the more stable −Sb−Te···Te−Sb− (or −Sb−Te−vacuum) stacking has larger gap than the −Ge−Te···Te−Ge− (or −Ge−Te−vacuum) stacking. It illustrates the distinct CTPs of these structures. In addition, we find non-uniform charge density of these structures at Γ point of valence-band maximum (VBM) and conduction band minimum (CBM), as shown in Fig. 6(e)-(l). It demonstrates that the boundary or vdW gap can tune the CTP. In the two-dimension models, the TTB of the −Sb−Te··· stacking repel both electron and hole in VBM and CBM, respectively, as shown in Fig. 6(e)-(f). The boundary with the −Ge−Te… stacking in Fig. 6(g)-(h) attracts the electron and repels hole in VBM and CBM, repspectively. In the bulk phases, we find that the vdW gap of −Sb−Te⋯Te−Sb− stacking attract the electron and repel the hole, respectively, as shown in Fig. 6(i)-(j). While the the vdW gap of −Ge−Te⋯Te−Ge− stacking repel the electron and repel the hole, respectively, as shown in Fig. 6(k)-(l).

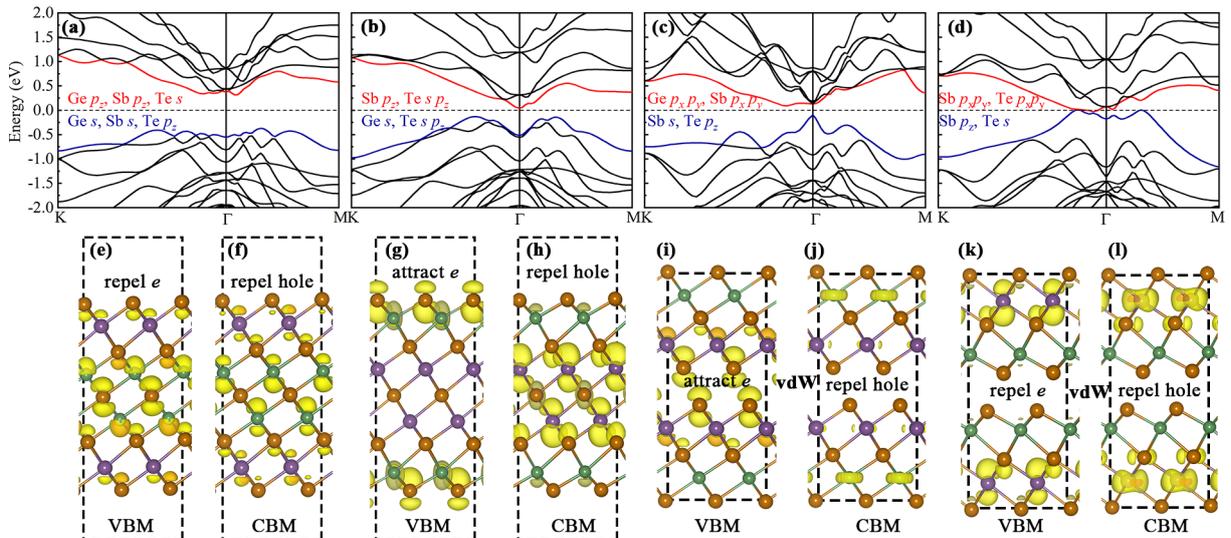

**Figure 6.** Electronic structures of two stackings in the two dimension and bulk phases. The band structures for two dimension **(a, b)** and bulk **(c, d)**. The Γ-point charge density of VBM and CBM for two dimension **(e-h)** and bulk **(i-l)**, where the 3D isosurface is 0.037eÅ$^{-3}$.

In short, we find that Γ-point charge distribution has two-dimension characteristic. Boundary thus hinders charge transport in the vertical direction and increases resistance. The high resistance in the phase-change working area enhances the heat efficiency against the heat loss in other circuit part. It is a key to achieve ultrafast speed by providing enough heat at short time, because actual circuit has threshold voltage, such as 2.5V in 40nm node. Previous work shown the decreased resistance as grain becomes larger during cycling operations in C−GST [25], where the beginning high resistance is mainly determined by boundary due to fewer inner vacancies in small grains.

**Conclusion**

In summary, we predict the optimal composition of various PCMs according to the MO theory. This quantum effect results in the distinct stability of TTB, which determines the boundary-related properties, such as distinct crystallization mechanisms, alterable inner vacancy concentration, and CTP. We predict that it exists a critical grain size of no inner vacancy, 4.7 nm for GST and 2.6 nm for ST, which is manifested by the absence of cubic-to-hexagonal transition in C−GST of small grains and minimized inner vacancy. These findings declare that we can improve device performance by the tunable boundary, such as via composition design or doping impurity. On the other hand, chalcogenide is used to fabricate lots of two-dimensional materials [55], which can be explained by forming stable boundary like in ST and GST reported here.

**Methods**

**First-principle calculations.** First-principle calculations are carried out using VASP package. [57] The Kohn-Sham equations are solved using the projector augmented wave (PAW) method [58] and Perdew−Burke−Ernzerhof with van der Waals correction (PBE-D3) [59,60] generalized gradient approximation (GGA) functional [61] with the kinetic energy cutoff of 300 eV, where the valence electrons are $3d^{10}4s^24p^2$ for Ge, $5s^25p^3$ for Sb, $5s^25p^4$ for Te, and $2s^22p^2$ for carbon. The convergence criterion is $1\times10^{-7}$ eV for electronic convergence and 0.05 eV/Å for force. A reasonably converged grid spacing of 0.02 Å$^{-1}$ is used. In order to calculate COHP and partial charge, the noncollinear calculation with the relativisitc SOC effects is not considered. First-principle molecular dynamics is carried out to obtain the nucleation rate of GST and GT, where we use a time step of 3 fs at 600 K with Parrinello-Rahman barostat and Langevin thermostat. The degree of crystallinity is judged by $q_6^{dot}$ parameter, the definition shown SI, which is a modified Steinhardt order parameter to describe the similarity of one atom with its surrounding atoms.

**COHP analysis.** High-precision calculations are carried out using VASP package, and then implemente the pCOHP bonding analyses using LOBSTER setup [62].

**Energy barrier calculation.** The energy barrier of boundary migration is calculated by the novel Stochastic Surface Walking (SSW) method [63] via smooth surface walking along softened random directions. We utilize recently developed SSW reaction pathway method to sample the grain-boundary-migration pathways. [64] The explicit transition states (TSs) of the pathways are located by the variable-cell double-ended surface walking (VC-DESW) method. [65] The calculation details are consistent with our previous work. [66]

**Experiment details.** GST, GeTe, and C−GST films (~100 nm in thickness) were deposited on $SiO_2/Si(110)$ substrate at room temperature by sputtering GST, GT, and C−GST alloy targets. The deposition proceeds with Ar at a flow rate of 50 SCCM (cubic centimeter per minute at standard temperature and pressure) with the $1\times10^{-5}$ Pa background pressure. By using of X-ray fluorescence spectrometer, the composition of GST and C−GST films were determined to be $Ge_2Sb_2Te_5$ and $C_{16}(Ge_2Sb_2Te_5)_{84}$. Thermal annealing processes were done in rapid thermal processing system. The annealed films were fabricated into TEM sample by using focus ion beam (FIB) in FEI Helios G3 system involving low pressure polishing process, and then cleaning in Gatan 691 PIPS. The STEM-HAADF analysis was carried out on JEM Grand ARM 300F microscope operated at 300 kV using STEM mode with probe corrector, the inner semi-angle of the detector is larger than 64 mrad. The grain size is analyzed by ACOM-TEM, which works by comparing the electron diffraction pattern collected from each scanning position with the simulated patterns calculated from a given template structure of all possible orientation maps [67]. ACOM-TEM analysis is performed by the ASTAR software from NanoMEGAS SPRL. More details are shown in SI.

**Acknowledgements**
Supported by the National Key Research and Development Program of China (2017YFA0206101, 2017YFB0701703, 2017YFA0206104, 2018YFB0407500, SQ2017YFGX020134), "Strategic Priority Research Program" of the Chinese Academy of Sciences (XDA09020402), National Integrate Circuit Research Program of China (2009ZX02023-003), National Natural Science Foundation of China (61874129, 61874178, 61504157, 61622408), Science and Technology Council of Shanghai (17DZ2291300, 18DZ2272800).

**Author contributions**
W.S., Q.T., and J.Z. contributed equally to this work. W.S. designed the research and carried out the calculations; W.S., Q.T., J.Z., Y.Z, and Y.C. implemented the TEM experiments; M.V. taken the ACOM-TEM characterization; all the authors analyzed the results and took the discussions; W.S. wrote the paper. The project was initiated by Zhitang Song. Authors to whom correspondence should be addressed: songwx@mail.sim.ac.cn, ycheng@ee.ecnu.edu.cn, and ztsong@mail.sim.ac.cn.

**Additional information**
The authors declare no competing financial interests.

# Supporting Information

Crystallization Mechanism Tuned Phase-Change Materials: Quantum Effect on Te-Terminated Boundary

*Wen-Xiong Song[1#]\*, Qiongyan Tang[#2], Jin Zhao[#1,3], Muriel Veron[4], Xilin Zhou[1], Yonghui Zheng[2], Daolin Cai[1], Yan Cheng[2]\*, Tianjiao Xin[1], Zhi-Pan Liu[5], and Zhitang Song[1]\**

[1]State Key Laboratory of Functional Materials for Informatics, Shanghai Institute of Microsystem and Information, Chinese Academy of Sciences, Shanghai 200050, China; [2]Key Laboratory of Polar Materials and Devices (MOE), Department of Electronics, East China Normal University, Shanghai 200241, China; [3]School of Physical Science and Technology, Shanghai Tech University, Shanghai 201210, China; [4]University Grenoble Alpes, CNRS, SIMAP, 38000 Grenoble, France; [5]Department of Chemistry, Fudan University, Shanghai 200433, China.

## Contents:

Part I. Mechanisms of ACOM-TEM technique;

Part II. A typical grain in GeTe at 180 °C;

Part III. Alterable inner vacancy concentration controlled by Te-terminated boundary;

Part IV. $q_6^{dot}$ parameter.

## Part I. Mechanisms of ACOM-TEM technique

ACOM-TEM works by comparing the electron diffraction pattern collected from each scanning position with the simulated patterns calculated from a given template structure of all possible orientation maps [1], as shown in Fig. S1(a)-(d), which was performed by the ASTAR software from NanoMEGAS SPRL. Here, we choose a metastable cubic phase of $Ge_2Sb_2Te_5$ (GST) as a template structure, a NaCl-type supercell with random vancancies distributed with basic lattice parameter $a$ = 6.07 Å. It is because amorphous GST at first crystallizes to a cubic phase at ~150 °C, which acts as a working crystalline phase in the PCRAM, and an undesired cubic-to-hex phase transition will occur at ~320 °C.

For each scanning position, the collected diffraction pattern matches with full set of templated patterns via structure rotation, and the best fit result is the most probable crystal orientation. A color map is constructed to represent the whole crystal orientations of each part in the sample, as shown in Fig. S1(e). To evaluate the fit quality, we define a correlation index $Q_i$ for the $i^{th}$ template structure as the following,

$$Q_i = \frac{\sum_{j=1}^{m} P(x_j,y_j)T_i(x_j,y_j)}{\sqrt{\sum_{j=1}^{m} P^2(x_j,y_j)}\sqrt{\sum_{j=1}^{m} T_i^2(x_j,y_j)}}, \quad (1)$$

where $P(x_j,y_j)$ and $T_i(x_j,y_j)$ are the intensity of measured diffraction patterns and the best fit patterns of the $i^{th}$ template structure (here only a cubic phase), respectively. $Q_i$ shows the degree of matching between the measured patterns and the fit patterns. The index map shows the the highest mathcing index at each position, as shown in Fig. S1(f).

TBs can be detected by ACOM-TEM, as shown in Fig. S1(e) and 1(f). With the help of the tilting-series scanning method, the trace analyses for twinning plane could be conducted successively and connectedly on the pole figures, as shown in Fig. S1(g). Figure S1(h) shows that the average grain size (in area) of GST is 20~40 nm at 250 °C.

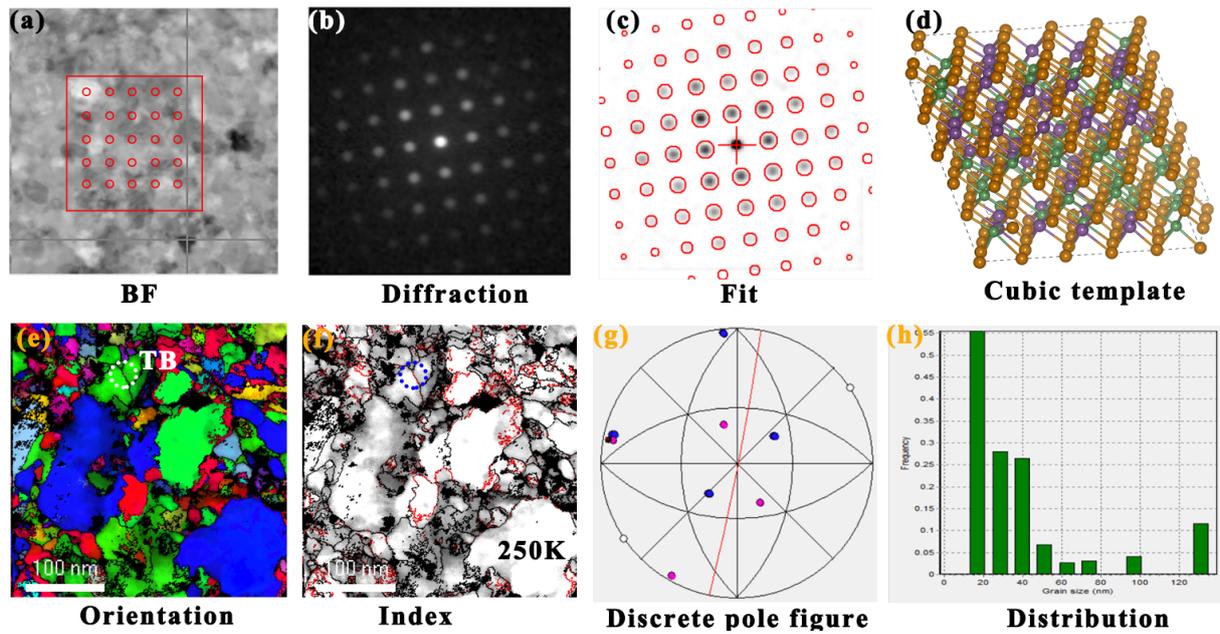

**Figure S1.** Working mechnism of ACOM-TEM and data interpretation. **(a)** Bright-field (BF) image of GST film annealed at 250 °C for half an hour, where the dot diameter of 1 nm represents the scan area, **(b)** diffraction pattern, **(c)** fitted pattern using the template structure of the cubic GST with random vacancies **(d)**. **(e)** Index maps where the red lines respent probable TBs, **(f)** orientation, **(g)** discrete pole figures for the circled TB in **(e)** and **(f)**, and **(h)** distribution of grain size.

**Part II. A typical grain in GeTe at 180 °C;**

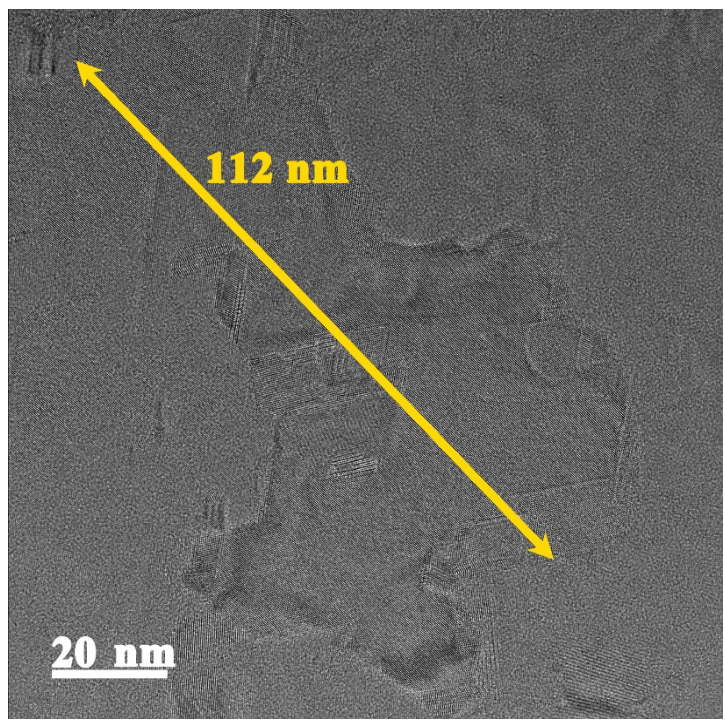

**Figure S2.** A typical grain in GeTe at 180 °C.

# Part III. Alterable inner vacancy concentration controlled by Te-terminated boundary

Applying a sphere grain model to investigate the Te-terminated (111) plane, the ratio of boundary atoms in the sphere grain is defined as,

$$I = \frac{4\pi R^2/S_a}{\frac{4}{3}\pi R^3/V_a}$$

$$= \frac{3V_a}{RS_a} \quad (2)$$

where $V_a$ is atomic volume (~27 Å$^3$ herein), $S_a$ is atomic area in the (111) plane (~15.6 Å$^2$ herein), and $R$ is grain size. Figure 4a shows the ratio of boundary atoms changes with grain size. It shows that small grain has much high ratio. For example, 26% atoms are in the boundary as grain is as small as 2 nm. The extra Te (or vacancy) content even exceeds the ratio of 11.11% in GST. We can predict that GST exists a critical grain size of 4.7 nm of no inner vacancy when the ratio is 11.11%. For ST, the critical size is 2.6 nm with the ratio of 20%. However, the ratio will reduce as grain grows. In order to keep the optimal local composition, the vacancy should be produced in the inner, which is manfisted by the vacancy ordering layers (VOL) near the boundary in Fig. 2(e); meanwhile, occurrence of mainly nine-atom layers meet the optimal chemical composition. For the sample annealing at low temperature, 180 °C for half an hour, short VOLs are produced in Fig. 4(h).

# PART IV. The $q_6^{dot}$ order parameter

Steinhardt's order parameter is defined in the following, [2]

$$Q_l = \left[ \frac{4\pi}{2l+1} \sum_{m=-l}^{l} \left| \sum_{b=1}^{N_b} Y_{lm}(\theta_b, \phi_b) \right|^2 \right]^{\frac{1}{2}} / N_b \quad (3)$$

where the spherical harmonics $Y_{lm}$ describes the local order of the centered atom surrounded by its nearest-neighbor atoms. The summation in equation (3) runs over all $N_b$ bonds in the first shell within a cutoff of 3.6 Å in this work.

However, Steinhardt's order parameter is not convenient for the condition of multiple crystallites instead of a single crystalline nucleus. A local version of $q_l$ can be defined for each atom in the following vector:

$$\boldsymbol{q_l}(i) = \begin{pmatrix} q_{l,l} \\ q_{l,l-1} \\ \cdots \\ q_{l,-l+1} \\ q_{l,-l} \end{pmatrix} = (q_{lm}(i))_{m=-l,l} \quad (4)$$

$$q_{lm}(i) = \frac{1}{N_i} \sum_{j \in \Omega_i} f_{ij} Y_{lm}(ij) \quad (5)$$

A radial cutoff function $f_{ij}$ is introduced to smooth the boundary:

$$f_{ij}(r) = \begin{cases} 1 & : r \leq r_1 \\ \frac{1}{2}\left\{\cos\left[\frac{\pi(r-r_1)}{(r_2-r_1)}\right] + 1\right\} & : r_1 < r \leq r_2 \\ 0 & : r > r_2 \end{cases} \quad (6)$$

The exponents in $f_{ij}$ were set to $r_1$ = 3.2 Å and $r_2$ = 3.6 Å.

The norm of $\boldsymbol{q_l}(i)$ is a local $Q_l$ version for an atom:

$$q_l(i) = \sqrt{\frac{4\pi}{2l+1}} \|\boldsymbol{q_l}(i)\| \quad (7)$$

The order parameter $q^{dot}$ is defined based on the bond order correlation $C_{ij}$ between neighboring atoms, first introduced by Frenkel and coworkers. [3]

$$C_{ij} = \frac{\boldsymbol{q_l}(i) \cdot \boldsymbol{q_l^*}(j)}{\|\boldsymbol{q_l}(i)\| \cdot \|\boldsymbol{q_l^*}(j)\|} \quad (8)$$

The order parameter $q_l^{dot}$ is the averaged sum of the bond order correlation $C_{ij}$, which is defined as the dot product of $q_l(i)$ and the complex conjugate of $q_l(j)$, divided by the rotationally invariant norm of the two vectors:

$$q_l^{dot}(i) = \frac{1}{N_i} \sum_{j \in \Omega_i} f_{ij} C_{ij} \quad (9)$$

The radial cutoff function $f_{ij}$ has the same form as above with the cutoff radius.